# The geography of references in elite articles:

# What countries contribute to the archives of knowledge

Lutz Bornmann,*[1] Caroline Wagner,[2] and Loet Leydesdorff [3]

[1] Division for Science and Innovation Studies, Administrative Headquarters of the Max Planck Society, Hofgartenstr. 8, 80539 Munich, Germany; bornmann@gv.mpg.de
[2] John Glenn College of Public Affairs, The Ohio State University, Columbus, Ohio, USA, 43210; wagner.911@osu.edu
[3] Amsterdam School of Communication Research (ASCoR), University of Amsterdam, PO Box 15793, 1001 NG Amsterdam, The Netherlands; loet@leydesdorff.net
*Corresponding author

# Abstract

This study is intended to find an answer for the question on which national "shoulders" the worldwide top-level research stands. Traditionally, national scientific standings are evaluated in terms of the number of citations to their papers. We raise a different question: instead of analyzing the citations to the countries' articles (the forward view), we examine referenced publications from specific countries cited in the most elite publications (the backward—citing—view). "Elite publications" are operationalized as the top-1% most-highly cited articles. Using the articles published during the years 2004 to 2013, we examine the research referenced in these works. Our results confirm the well-known fact that China has emerged to become a major player in science. However, China still belongs to the low contributors when countries are ranked as contributors to the cited references in top-1% articles. Using this perspective, the results do not point to a decreasing trend for the USA; in fact, the USA exceeds expectations (compared to its publication share) in terms of contributions to cited references in the top-1% articles. Switzerland, Sweden, and the Netherlands also are shown at the top of the list. However, the results for Germany are lower than statistically expected.

# 1 Introduction

The invention of the *Science Citation Index* in the 1960s was welcomed by citation analyst as well as historians and philosophers of science as an opportunity to make studies of science empirically informed (e.g., Cole & Cole, 1973; Elkana *et al*., 1978; Price, 1951 and 1965; cf. Gingras, 2016). As it developed, however, citation analysis became the common tool that split the two fields into holding different views on the reference as "tool." Citation analysts are interested in counting "times cited" to aid evaluation by measuring quality or impact. References in the other direction of "citing" came to be seen as a tool to indicate "revolutions and reconstructions in the philosophy of science" (Hesse, 1980). Wouters (1999) noted the inversion of the cited/citing matrix by the database owner as added value which shapes the field of citation analysis as an analytically different domain.

Kuhn's (1962) "disciplinary matrix," for example, can first be considered as a citation matrix, but Kuhn was not interested in breakthroughs and milestones as highly-cited papers, but in referencing: what is referenced when scholars cite (Kuhn, 1984)? Can a "disciplinary matrix" be indicated? The inversion is known in citation analysis as the difference between co-citation (Small, 1983; Marshakova, 1973) and bibliographic coupling (Kessler, 1963). By citing, a scholar reconstructs the intellectual context of one's knowledge claim (Fujigaki, 1998); being cited, however, is rewarding in terms of the sciences as a social enterprise since it provides reputation. Whitley (1984) submitted that from a sociological perspective, the sciences can be considered as reputationally-controlled work organizations. The intellectual organization of the sciences, however, evolves in—often anonymized—texts that are cleaned in a process of validation from the contingencies of the context of discovery (Popper, [1935] 1959).

In this study, we focus on the question of what is referenced in the worldwide top-level research in terms of national contributions. Which countries have provided the intellectual context of research? The intangible asset of reputation is an asset for nations, and this is reflected, in part, by referencing behavior. "Esteem" (as contrasted to fame) is earned over a long period of time and can be retained as a halo effect even after quality has reduced. Reputation, which is the "product of years of demonstrated superior competence" (Hall, 1992, p. 143), attracts resources such as funds and flows of talent. Thus, identifying the geographical shares of references point to those places that retain reputation.

We were inspired to take this approach by bibliometric analyses for the Science and Engineering Indicators report of the US National Science Foundation (National Science Board, 2012). Mervis (2012) highlighted in this report the shares of specific countries and transnational units such as the European Union in the references of 1% most-frequently cited papers. These shares were size-normalized by using the countries' numbers of published papers as the baseline. We consider this approach as groundbreaking for the question on which "shoulders" the worldwide top-level research stand (Bornmann, de Moya-Anegón, & Leydesdorff, 2010; Merton, 1965): which literature is incorporated in the archive that is cited at the top of the pyramid?

Operationally, we include the 21 countries in this study which published more than 1% of the articles worldwide during the period under study. Using the years 2004 to 2013, we investigate a substantially longer and more recent time period. We focus on countries as units of analysis because national systems represent underlying cultural, social, economic, and political models; national governments seek to encourage knowledge creation, diffusion, and exploitation; most basic research is paid for by public funds. In addition to activities at the research front, participation, prestige, and the build-up of intellectual capital are longer-term objectives of national policies. Prestige can be considered as generalized from performance (Brewer, Gates, & Goldman, 2001).

For example, the label "Made in Italy" has a value that can be compared with "Made in China" in terms of assumption about quality; the 'capital' attached to "Made in Italy" has been built up over time and with attention to maintaining quality. Prestige in science attracts foreign students and collaborators who can contribute to the vitality of a system (see, e.g., Adams, 2010; Adams, Pendlebury, & Stembridge, 2013; Leydesdorff & Wagner, 2009; Marshall & Travis, 2011; National Science Board, 2012). We operate on the basis of the expectation that top-level researchers are scanning internationally for knowledge, and that they seek to connect with equally or more elite collaborators to maximize their own reputation. Thus, references no longer reflect a local bias (Kaplan, 1965) but rather reflect an underlying dynamic of preferential attachment (Wagner & Leydesdorff, 2005).

For this study, we consider publications as investments in intellectual capital, reflecting the investments in maintaining quality reflected in one can also ask for a return on investment (King, 2004). We address this question by distinguishing between domestically and internationally (co-)authored publications. To which extent are authors of top-1% publications citing from national sources or internationally?

In other work, we showed that, in terms of field-normalized performance for the top-1% and top-10% most-frequently cited publications, the USA held the lead in the 2000s, but with increasing citations going to EU28 nations, which had also increased their share of articles in the top 10% highly cited articles (Leydesdorff, Wagner, & Bornmann, 2014). Several smaller European nations—Switzerland, Denmark, Sweden, and the Netherlands—had surpassed the USA in percentage share of highly cited articles. Writing in *Science*, Mervis (2012) showed that Asian scientists increasingly tend to cite other Asian articles. He also showed that the USA remains the leader in producing the top-1% most highly cited articles, but that the European Union, China, and other Asian countries are also growing in shares of the top 1% most highly cited articles. However, Mervis' (2012) report was limited to only five countries or aggregates of countries (e.g. Asia-8).

In this study, we expand upon Mervis' (2012) report and use the cited references to view national contributions to the archives of knowledge.

## 2   Data and Methods

We focus on the 21 countries that contributed 1% or more of all published material with the document type "article" in Web of Science (WoS) between 2004 and 2013 (Table 1). These 21 countries published 86% of all articles indexed in these years (across all subject categories). Using this threshold, most countries worldwide with a substantial contribution to the archive are included; however, small-sized but potentially top-performing countries in terms of relative citation impact such as Denmark are excluded (see Leydesdorff, Wagner, & Bornmann, 2014).

From the set of articles published between 2004 and 2013 with one of these countries in the address field, we created a dataset containing the top-1% most highly cited research worldwide. We call these papers "elite" articles. "Elite" articles are those articles which belong to the 1% most-frequently cited papers in the corresponding WoS subject categories and publication years. From this dataset, we collected all the cited references. This resulted in a subset of the data that we further cleaned in three ways, removing about 40% of the material. 1) We eliminated books, books chapters, news media, conference papers, notes, letters, dissertations, and other types of sources, since it is not always possible to quantify country-level contributions to these materials and referencing patterns differ. Thus, we included only references with the document type "article". 2) We removed articles lacking country information in the address lines—otherwise we could not link knowledge contributions to countries, which is the point of the analysis. 3) We eliminated any referenced articles dated prior to 1980 because the database does not contain reliable address information prior to this date.

The final data contains a mix of articles, with some listing one and others listing more than one country in the address line. If an article has multiple country names in the address lines, we calculated fractional counts of contributions to articles based upon the numbers of countries. The counting process involved allocating a fractional share based upon the number of countries represented in an article's address lines. The count goes to countries—not to authors: If multiple authors are listed from the same country, the count is still "one" for this country. If two authors from two different countries are listed on the article, the article is assigned to each country with a value of 0.5; for three countries, the value is .33, and so on.

Table 1. Twenty-one countries with the largest shares of all articles indexed in WoS between 2004 and 2013. Only countries with more than 1% of the fractionally-counted articles are listed, in decreasing order of the percentage share of articles

| Country | Absolute numbers of articles (full counting) | Absolute numbers of articles (fractional counting) | Percentages of share (fractionally counted) |
|---|---|---|---|
| USA | 3,168,104 | 2,634,682.58 | 24.02 |
| China | 1,235,872 | 1,080,633.73 | 9.85 |
| Japan | 749,737 | 642,650.14 | 5.86 |
| UK | 852,450 | 610,480.36 | 5.57 |
| Germany | 825,301 | 588,873.30 | 5.37 |
| France | 591,754 | 415,985.78 | 3.79 |
| Canada | 499,266 | 368,465.44 | 3.36 |
| Italy | 469,169 | 348,992.92 | 3.18 |
| India | 367,526 | 324,676.10 | 2.96 |
| South Korea | 364,148 | 309,488.83 | 2.82 |
| Spain | 403,738 | 302,138.84 | 2.75 |
| Australia | 354,499 | 260,940.89 | 2.38 |
| Brazil | 276,151 | 233,401.27 | 2.13 |
| Russian Federation | 264,695 | 212,192.69 | 1.93 |
| Taiwan | 219,447 | 192,525.35 | 1.76 |
| Netherlands | 276,224 | 186,951.06 | 1.70 |
| Turkey | 197,794 | 178,076.95 | 1.62 |
| Poland | 178,423 | 140,765.21 | 1.28 |
| Sweden | 189,527 | 126,237.57 | 1.15 |
| Switzerland | 201,586 | 122,881.07 | 1.12 |
| Iran | 137,972 | 122,174.78 | 1.11 |

Table 1 shows the countries with the largest shares of all articles between 2004 and 2013. As expected, the United States (USA) is at the top of the list of countries, followed by China, Japan, the UK, and Germany as the countries contributing the largest numbers of articles. (The European Union is not considered as a single unit in this analysis.) China appears second in the total numbers of articles. However, it did not begin the decade in this second position, but grew much more rapidly than other countries to finally claim the second spot (Zhou & Leydesdorff, 2008).

## 3     Exploring international contributions

Table 2 lists countries in the same order as Table 1; it shows the number of references per country and their respective shares of contributions to elite publications. For example, China published 9.85% of the worldwide articles (Table 1), but it contributed only 4.24% of cited references in the top-level research papers. The opposite is seen for the USA: it contributes 24.02% of worldwide articles (Table 1) and 44.1% of the references in top-1% (Table 2). This large share for the USA led us to consider weighting the data, since a factor accounting for the differences may be the publication volume of the various countries (see here Harzing & Giroud, 2014). The more articles a country has published, the more citations can *ceteris paribus* be expected.

Table 2. Country counts from cited references in elite publications. Sorting order matches Table 1.

| Country | Absolute numbers (full counting) | Absolute numbers (fractional counting) | Percentage share (fractionally counted) |
|---|---|---|---|
| USA | 1,690,279 | 1,403,550.91 | 44.10 |
| China | 177,660 | 134,969.53 | 4.24 |
| Japan | 194,537 | 150,026.31 | 4.71 |
| UK | 378,794 | 247,783.00 | 7.79 |
| Germany | 299,068 | 184,287.05 | 5.79 |
| France | 204,445 | 121,722.31 | 3.82 |
| Canada | 199,270 | 128,511.17 | 4.04 |
| Italy | 136,224 | 79,005.36 | 2.48 |
| India | 35,543 | 26,136.91 | 0.82 |
| South Korea | 53,244 | 37,210.09 | 1.17 |
| Spain | 91,964 | 54,105.14 | 1.70 |
| Australia | 122,767 | 77,797.85 | 2.44 |
| Brazil | 27,888 | 15,223.68 | 0.48 |
| Russian Federation | 30,163 | 13,051.16 | 0.41 |
| Taiwan | 30,409 | 22,615.22 | 0.71 |
| Netherlands | 132,073 | 78,608.76 | 2.47 |
| Turkey | 16,041 | 12,018.90 | 0.38 |
| Poland | 24,378 | 11,146.73 | 0.35 |
| Sweden | 81,830 | 47,488.45 | 1.49 |
| Switzerland | 102,371 | 56,292.08 | 1.77 |
| Iran | 8,092 | 6,366.80 | 0.20 |

Figure 1 shows the same data as Table 2, but broken down by year of publication. As expected, the USA's contribution is decreasing (Wagner, 2011), whereas China's contribution is increasing over the years. The USA share of cited references in top-1% articles dropped by approximately nine percentage points—more than the drop in publication volume (Leydesdorff et al., 2014). However, China's share increased by 5.7 percentage points, which is proportional to its gain in shares of publications. We included 95% confidence intervals in Figure 1 for China and the UK as interval estimates indicating the accuracy of our point estimates (the percentages) (Williams & Bornmann, 2016). The absence of overlap of the 95% confidence intervals for the UK and China shows that the lead of the UK versus China is (still) statistically significant at the end of the period (Cumming, 2012; Cumming & Finch, 2005). The shares of the other countries are more or less constant over the years.

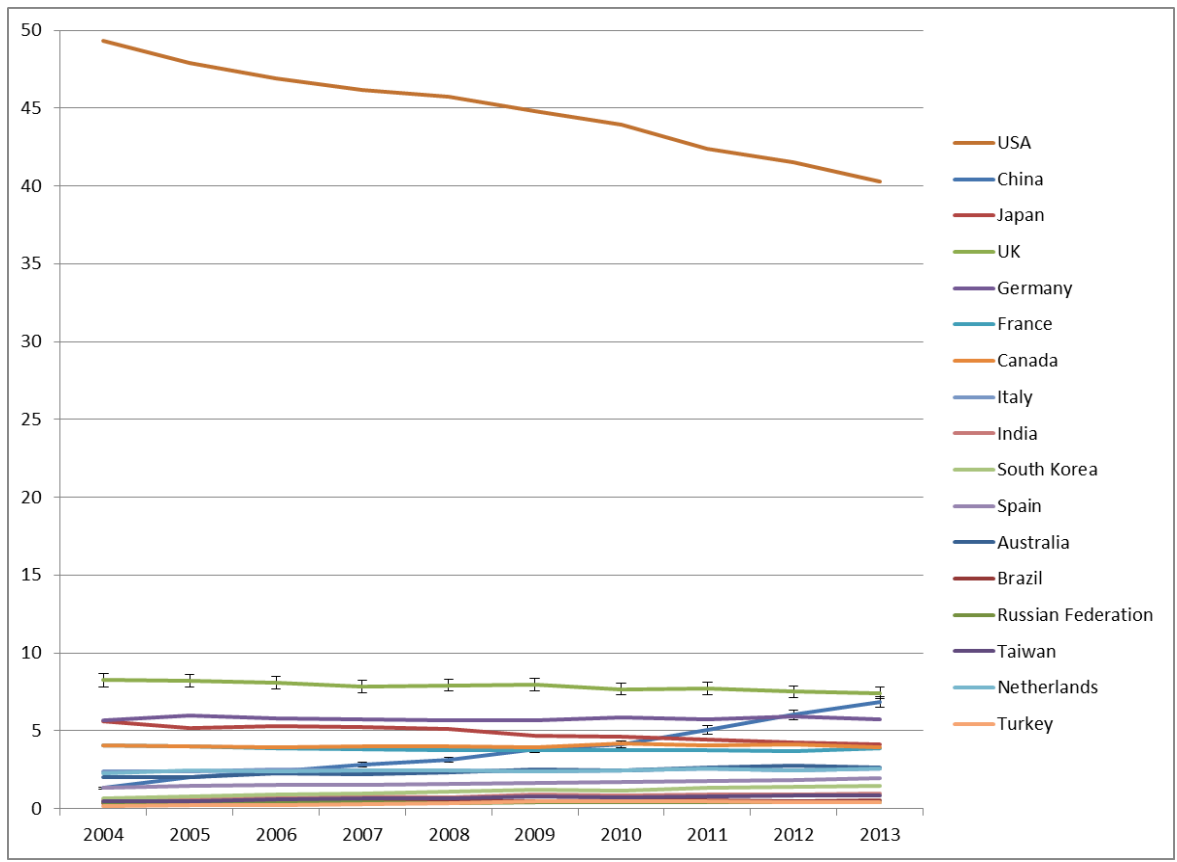

Figure 1. Countries' shares of references cited in the elite articles between 2004 and 2013 (articles belonging to the 1% most frequently cited articles, fractionally counted). 95% confidence intervals are added to the UK's and China's shares.

In Table 3, we report the ratio of cited references in the elite articles to citing articles for each country. Assuming that many papers reach their citation peak in the third year after publication, we use the ratio of cited references in year *t* and published articles from year *t*-3. This ratio reveals whether a country received more citations than expected on the basis of the number of published articles. The findings show that the USA has an average ratio of 1.7 (cited references) and (citing articles) during this period. Thus, the USA contributed much more to the archive than can be expected on the basis of its publication volume. Maisonobe, Milard, Jégou, Eckert, and Grossetti (2017) report on similar results for the US. Besides the US, Switzerland, the Netherlands, the UK, and Sweden had higher-than-expected citedness compared to publication volume in this study.

Table 3. Mean ratios and standard deviations of shares (cited references versus citing articles) across 10 years as well as the difference between the ratios in the last (2013/2010) and first years (2004/2001) (sorted by the means).

| Country | Mean | Rank order | Standard deviation | Rank order | Last year - first year | Rank order |
|---|---|---|---|---|---|---|
| USA | 1.71 | 1 | 0.03 | 17 | 0.03 | 17 |
| Switzerland | 1.53 | 2 | 0.08 | 5 | 0.21 | 6 |
| Netherlands | 1.39 | 3 | 0.09 | 3 | 0.28 | 2 |
| UK | 1.29 | 4 | 0.08 | 6 | 0.21 | 5 |
| Sweden | 1.20 | 5 | 0.06 | 8 | 0.12 | 11 |
| Canada | 1.17 | 6 | 0.03 | 16 | -0.02 | 20 |
| Australia | 1.04 | 7 | 0.09 | 4 | 0.18 | 7 |
| Germany | 0.99 | 8 | 0.10 | 2 | 0.27 | 3 |
| France | 0.91 | 9 | 0.07 | 7 | 0.22 | 4 |
| Italy | 0.75 | 10 | 0.04 | 12 | 0.12 | 10 |
| Japan | 0.69 | 11 | 0.05 | 9 | 0.11 | 12 |
| Spain | 0.62 | 12 | 0.04 | 11 | 0.16 | 8 |
| China | 0.50 | 13 | 0.10 | 1 | 0.34 | 1 |
| South Korea | 0.45 | 14 | 0.04 | 13 | 0.12 | 9 |
| Taiwan | 0.42 | 15 | 0.04 | 14 | 0.10 | 13 |
| India | 0.30 | 16 | 0.03 | 19 | 0.03 | 16 |
| Poland | 0.28 | 17 | 0.02 | 21 | 0.01 | 18 |
| Iran | 0.27 | 18 | 0.05 | 10 | 0.07 | 14 |
| Brazil | 0.26 | 19 | 0.03 | 18 | -0.03 | 21 |
| Turkey | 0.25 | 20 | 0.03 | 15 | -0.01 | 19 |
| Russian Federation | 0.19 | 21 | 0.02 | 20 | 0.07 | 15 |

Note. The rank positions were calculated on the base of mean values, which are not rounded to two decimal places.

Table 3 shows the mean ratio and standard deviation of shares (cited references versus published articles) that were calculated across 10 years, as well as the difference between the ratios in the last (2013/2010) and first years (2004/2001). By using the mean values, the countries can be categorized into three groups indicated as grey-shaded areas in the table: high, average, and low performers. The high performers have a ratio of at least 1.2, which means that they received substantially more citations in the elite publications than would be expected by publication volume (on average across the years). The average performers approximately meet the expectations (values between 0.8 and 1.19). The low performers fall significantly below the expectations (below 0.8) based upon volume. Table 3 shows that China is still on a low performance level (rank position for the mean is 13), but grows more quickly than the other countries across the years (rank position for the standard deviation is 1). The differences between the last (2013/2010) and first years (2004/2001) in the table point out that China has a high increase in its ratio (0.34), but other countries have a positive showing as well (e.g., the Netherlands=0.28 and Germany=0.27).

Figure 2 shows the developments of the countries' ratios of cited references versus citing articles over time. The countries are categorized into three groups of performers as per Table 3: the top box includes USA, UK, the Netherlands, Sweden, and Switzerland—the high performers. Since the group consists of both large and small countries (in terms of published articles), the size-normalization used above seems to function properly. As Figure 2a reveals, the USA performs at the highest level across all years. In recent years, Switzerland has reached a very high level, too. The UK and the Netherlands show an increasing trend.

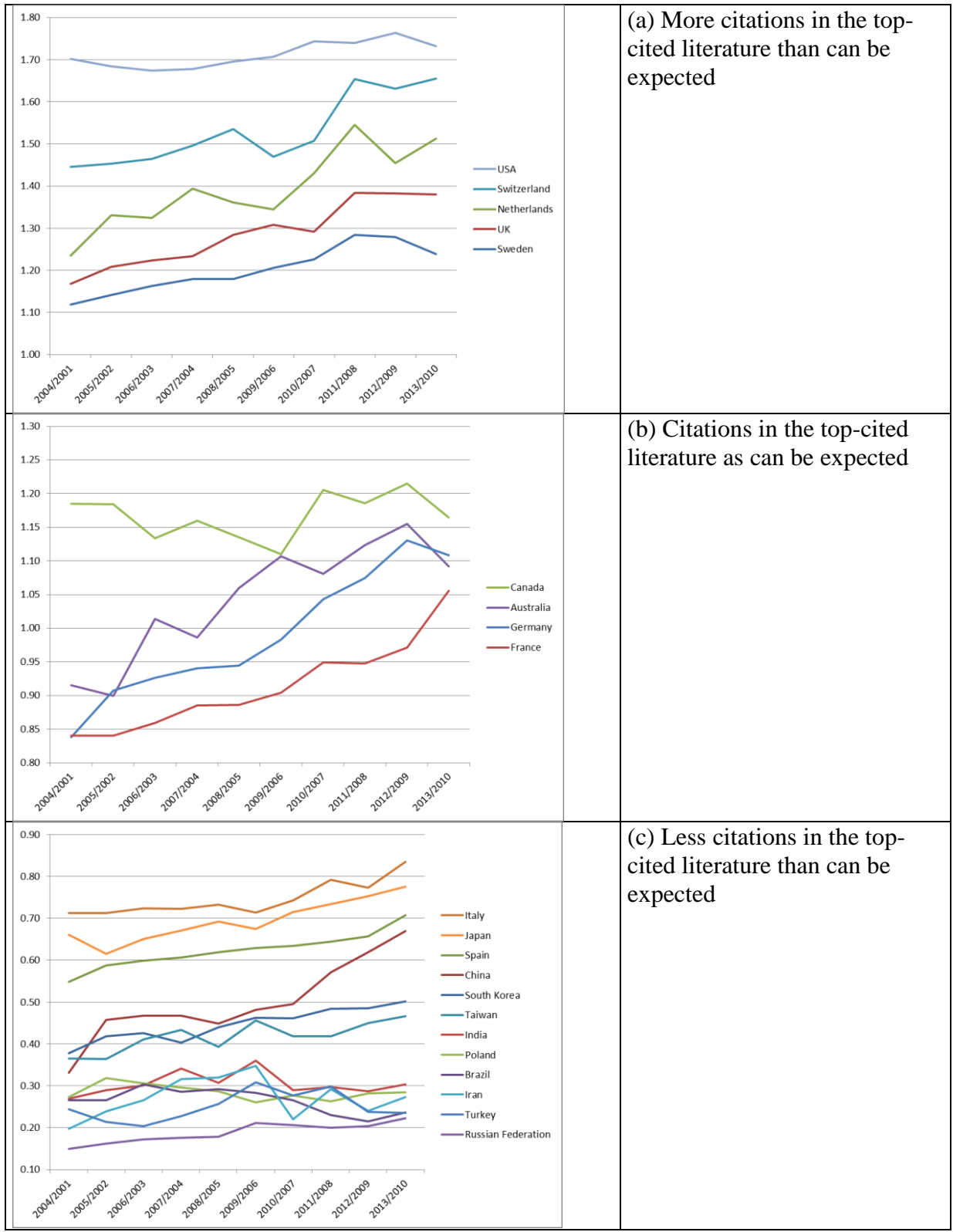

Figure 2. Ratios of shares based on (1) cited articles: countries' shares of references cited in the top-level research (fractionally counted). (2) Published articles: countries' shares of articles published between 2001 and 2010 (fractionally counted). The countries are categorized as high (a), average (b), and low (c) performers (see Table 3).

The average performing group in Figure 2b consists of four countries with ratios around 1 across the years (Germany, France, Canada, and Australia). With the exception of Canada, these countries show an increasing trend. The largest number of countries (n=12), however, are categorized as belonging to the low-performing group, shown in Figure 2c. Several countries show an upward trend, notably China, which shows an upward turn beginning in 2005. Other countries also show upwards trends, including Italy (since 2009/2006), which performs in the average range by the end of the decade.

# 4    Exploring Domestic Contributions

As a second research question, we are able to distinguish between each country's contribution to the archive of the international literature versus the domestic return on investment: how much does a country itself profit from this longer-term incorporation of its contribution to the elite literature? We operationalize this domestic effect by normalizing the the country-level contributions to the cited references against the set of elite articles published by the country without considering internationally (co)authored articles. Thus, we would like to focus on the countries stand-alone strength by including only domestic (cited) references and (citing) top-1% articles in the analysis.

The result is shown in Figure 3, which is rather similar to Figure 2. However, this similarity that is telling. The focus on domestic articles in Figure 3 supports the previous results and suggests that the global-level contributions reflect the national efforts and strengths. However, two interesting differences are visible for countries in the top group. (1) When the analysis is limited to domestic articles, the excellent performance of the USA becomes *more* pronounced, suggesting that US authors of top-1% articles are more likely to cite one another than authors from abroad. (2) Since there is a larger gap between the USA and the other countries, the contribution of the USA seems to reflect its domestic strength.

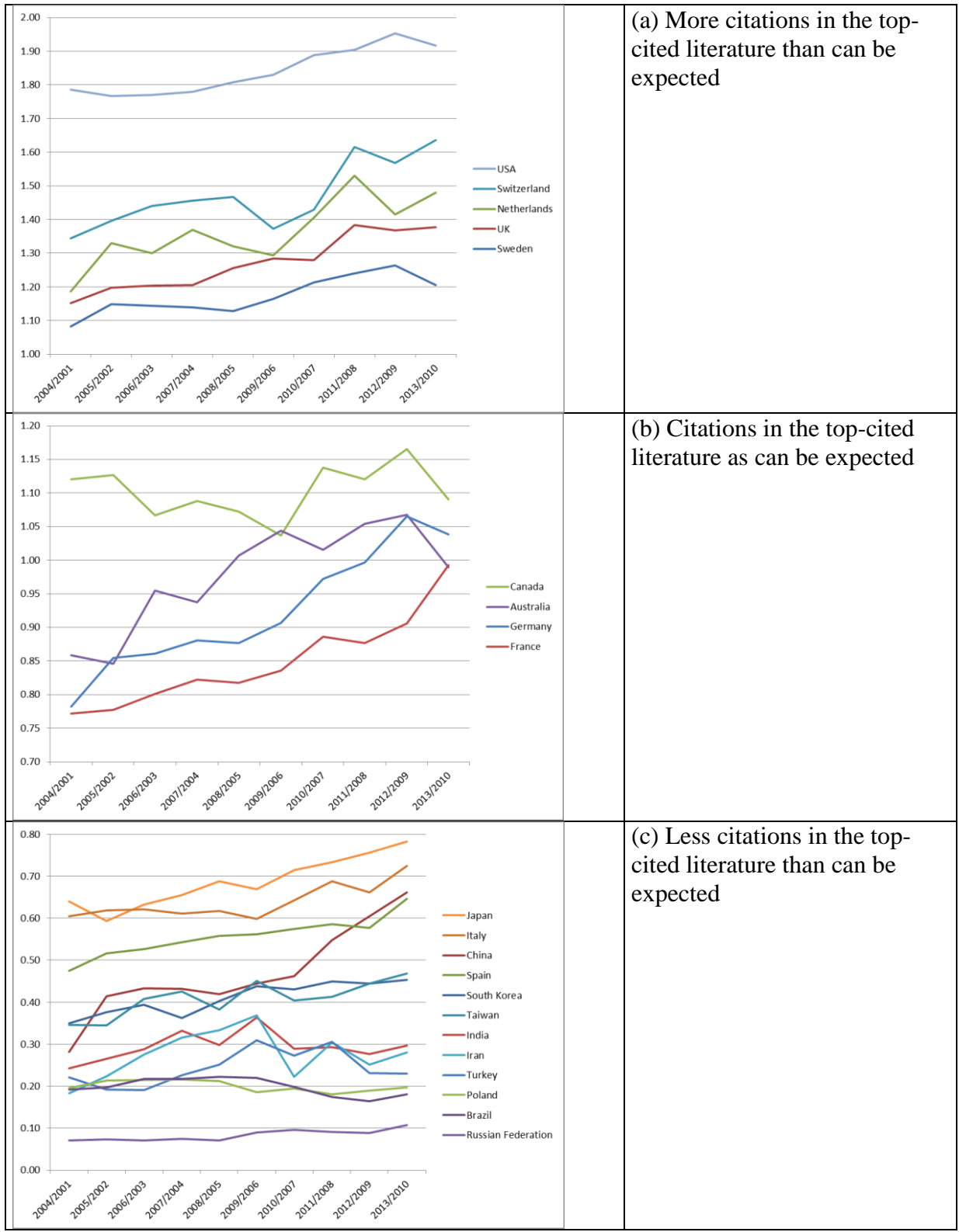

Figure 3. Moving averages of ratios of shares between (1) countries' shares of references cited in elite articles (fractionally counted) and (2) countries' shares of articles published between 2001 and 2010 (fractionally counted). The countries are categorized as high (a), average (b), and low (c) performers (as listed in Table 3).

The more articles from a country receive citations in its own top-level research papers, the more a country can be considered to have contributed to its own knowledge basis and national strength. With a high share of cited references in the national top-level research, the investments in science seem to be used in the most efficient way from the perspective of a national government. Since the shares of cited references cited in the national top-level research papers is not only dependent on the quality of research, but also on the publication volume, the shares of cited references should be size normalized.

The comparison of shares in cited references and citing articles can therefore reveal how a country's scientific base benefits from the nation's investment in research or whether these investments spill-over the national borders. In order to make this assessment, the domestic shares of cited articles in each country (subset of dataset 2) are contrasted to the shares of published articles. If the ratio is larger than 1, a benefit to the country can be inferred, or, a "gain" of investment from a national perspective (Yan, Ding, Cronin, & Leydesdorff, 2013).

Table 4. Ratios of two shares: (1) share of the number of cited references in the country's articles and the number of that part of cited references which were published by the country itself. (2) Country's shares of articles published between 2001 and 2010 (fractionally counted).

| Year | USA | Netherlands | UK | Switzerland | Sweden |
|---|---|---|---|---|---|
| 2004/2001 | 2.34 | 7.22 | 3.60 | 9.87 | 9.97 |
| 2005/2002 | 2.34 | 8.00 | 3.46 | 10.10 | 8.23 |
| 2006/2003 | 2.38 | 7.76 | 3.71 | 10.78 | 8.97 |
| 2007/2004 | 2.41 | 8.08 | 3.64 | 9.94 | 9.33 |
| 2008/2005 | 2.48 | 7.48 | 3.66 | 9.98 | 10.09 |
| 2009/2006 | 2.50 | 7.60 | 3.86 | 8.77 | 9.24 |
| 2010/2007 | 2.56 | 6.50 | 3.34 | 7.25 | 8.18 |
| 2011/2008 | 2.59 | 7.03 | 3.60 | 8.63 | 9.30 |
| 2012/2009 | 2.70 | 6.63 | 3.78 | 7.87 | 8.36 |
| 2013/2010 | 2.67 | 7.13 | 3.66 | 7.48 | 7.58 |

Table 4 shows the result of this domestic analysis for the five countries identified as top-performers: (1) the share of cited references articles in a country's top-level research papers that were published by the country itself among all cited articles from that country. (2) The share of articles published by the country among all published articles. For example, the share of domestically cited references among the total of cited references in the top-level research articles from 2005 is 66.71% for the USA. The share of US articles among all articles worldwide in 2002 is 28.47%. The resulting ratio is 2.34.

Return on investment at the national level far outperforms expectation for the Netherlands, Sweden, and Switzerland: domestic articles by authors in these countries are significantly more frequently cited in the top-level research than one would expect from their respective shares of published papers. However, the trend for Switzerland is decreasing. Thus, Switzerland's research becomes less important for its own top-level research papers across the years.

## 5  Discussion

When Garfield (1972; Garfield & Sher, 1963) introduced the JIF as a two-year moving average of journal citations, he based this decision on Martyn & Gilchrist's (1968) evaluation of British scientific journals (p. 476, n. 30). However, these authors had focused mainly on journals in molecular biology and biochemistry. In these fields more than 25% of the citations is provided in the first two years after publication (Garfield, 2003, p. 366). The JIF thus discounts the effects of "citation classics" (Bensman, 2007). The focus is on the rapidly-moving research front.

The sciences differ in terms of how relevant this "research front" is for the development of a field (Price, 1970). Short-term citation at a research front can be distinguished from longer term processes of incorporation and codification of knowledge claims into bodies of knowledge. Citation classics may not be highly cited in the first few years (Leydesdorff, Bornmann, Comins, & Milojević, 2016), but peak later.

For example, the *American Journal of Sociology* (AJS) and *American Sociological Review* (ASR)—two leading sociology journals—have cited and citing half-life times of more than ten years. In other words, more than half of the citations of these journals are from issues published more than ten years ago, and more than half of the references are to publications older than ten years. Coleman's (1988) study entitled "Social Capital in the Creation of Human-Capital," for example, became a most highly-cited paper almost two decades after its publication (cf. van Raan, 2004). Citation classics are not decaying as the citation curves of "normal science."

Focusing on short-term impact, one measures, from this perspective, not quality but variation. The selection mechanisms in the citation networks (Kuhn's (1962) "disciplinary matrix") can be expected to develop much more slowly (Hesse, 1980). By choosing a ten-year citation window and only top-1% citing papers, we conjecture that the best scientific papers are used as sources and thus still referenced by the top papers ten years later. Our approach is "citing" as different from "cited:" using the top-1% elite papers—normalized for fields of science—we can retrieve co-reference patterns (bibliographic coupling; Kessler, 1963) for previous literature.

In a so-called "linked" citation database, one can retrieve bibliometric characteristics of the referenced literature as backward citation as easily as forward citation rates. In this study, we focus on the geographical origins of the knowledge contributions by elaborating on a design reported by Mervis (2012). We use the addresses in the bylines of the "citation classics" to attribute them to countries. In order to avoid noise by unprecise referencing in the margins of scientific developments, we focus on the top-1% elite of scientific papers (normalized for fields). We assume that these authors have worked very carefully on their papers, including highly precise and selective referencing.

Based on previous studies, we expected to see a wider field of contributing countries to these top level papers, but our results show that the US science system is very strong in contributing to the global knowledge pool, and is heavily relied upon as the source of knowledge by both US and non-US authors. The USA remains the center of science in terms of the citation practices of other scientists who are seeking to advance research. Although the USA may be losing ground in science in other respects ( Leydesdorff & Wagner, 2009; National Science Board, 2004, 2010), this analysis suggests that American authors contribute more to the archives of elite global science than would be indicated by its number of published articles. The size-normalized contribution reveals that the USA has gained ground over other countries instead of losing it. This becomes especially visible if the analysis focusses on domestic publications.

Rodríguez-Navarro and Narin (2017) who compared the European Union with the USA have published similar results. Their analyses of publications belonging to the 1% most frequently cited demonstrate the ongoing dominance of the US in science. On the other side, the analyses of Rodríguez-Navarro and Narin (2017) also reveal the deficiencies of the European Union in the top-level segment. The European Union is, however, a very heterogeneous set of countries in terms of scientific performance, which should not be analyzed in the aggregate. The results of our analyses show that Switzerland, Sweden, and the Netherlands have high contributions to the cited references used in the top-1% elite articles, as has the USA—a finding similar to results of studies showing these nations as garnering more citations to their work. The surplus-capacity of these countries, measured by the shares of cited references and citing articles, is very high, putting them as among the most productive (and probably effective) of elite science in the world. These high performing countries exist alongside many other European countries with comparably medium or low performance.

Somewhat to our surprise, our results show that Germany does not belong to the top-performing group of countries. This result differs from the results of impact-oriented studies, which have demonstrated high performance for Germany in recent years (e.g., Bornmann &

Leydesdorff, 2013) and historically (King, 2004). The results also differ from the results which have been recently presented by Abbott (2017) in a *Nature* comment. The comment is entitled as "the secret to Germany's scientific excellence" and the presented numbers (e.g., the field-weighted citation impact) "tell a positive story for science" (p. 22). While German articles generally show strong short-term citation impact, Germany's long-term contributions to the elite literature is not as strong. It has been argued that German scientists are not as likely to publish in high impact journals as others (Murmann, 2013), which may reduce visibility of German research. It appears that German scientists spend less time on collaboration than peers in other top nations (Perkmann et al., 2013) and this may reduce the opportunity to contribute to cutting-edge problems. However, one can also argue that Germany is successful in optimizing profit from its investments by maintaining a national publication system. This question of Germany's efficiency requires further study.

The analysis further confirms that China has emerged as a major player in science, at least in terms of numbers of articles (Fu, Chuang, Wang, & Ho, 2011; Zhou & Leydesdorff, 2006, 2008). We expected to find that others increasingly draw upon China's science; however, our analysis suggests that China's contributions to the literature are still less relevant for elite publications. There may be many reasons for this beyond quality of research—social networks and language capabilities play roles in the dissemination of scientific knowledge, and these may remain obstacles for many Chinese scientists (Cao, Li, Li, & Liu, 2013; van Leeuwen, Moed, Tijssen, Visser, & van Raan, 2001).

Two other countries deserve mention. This is the poor showing of the Russian Federation and Japan. These two nations have declined in science from former leading positions. King (2004) showed Japan as the fourth strongest country in the world using data from 1997 – 2001 based upon the top 1% of highly cited publications. In this study, we show that Japan is contributing to references in elite research articles below expectation. This may be due, in part, to the fact that Japan is among the least internationalized nations in percentage terms (Wagner & Jonkers, forthcoming). Within Japanese culture, it is important to publish findings in Japanese language journals, which may reduce the dissemination of knowledge.

The position of the Russian Federation is more difficult to interpret because historical continuities have been disrupted by the break-up of the Soviet Union. Even so, in the 1990s, Russia was an average performing country, counted by King (2004) as close to Finland and Denmark in producing science and claiming citations. Unlike Japan—which has continued to fund R&D at a high rate—Russian investment in R&D has declined. As in Japan, publishing in the national language is important in Russia for one's reputation; but this orientation may hinder international visibility.

# Acknowledgements

The bibliometric data used in this paper are from an in-house database developed and maintained by the Max Planck Digital Library (MPDL, Munich) and derived from the Science Citation Index Expanded (SCI-E), Social Sciences Citation Index (SSCI), and Arts and Humanities Citation Index (AHCI) prepared by Clarivate Analytics, formerly the IP & Science business of Thomson Reuters.